\documentclass{article}

\usepackage{PRIMEarxiv}

\usepackage[utf8]{inputenc} 
\usepackage[T1]{fontenc}    
\usepackage{hyperref}       
\usepackage{url}            
\usepackage{booktabs}       
\usepackage{amsfonts}       
\usepackage{nicefrac}       
\usepackage{microtype} 
\usepackage{fancyhdr}       
\usepackage{graphicx}       
\graphicspath{{../Figures/}}     

\usepackage{float}
\usepackage{subfigure}
\usepackage{graphicx}%
\usepackage{multirow}%
\usepackage{amsmath,amssymb}%
\usepackage{amsthm}%
\usepackage{mathrsfs}%
\usepackage[title]{appendix}%
\usepackage{xcolor}%
\usepackage{textcomp}%
\usepackage{manyfoot}%
\usepackage{natbib}
\usepackage{algorithm2e}

\title{
Simple yet effective: a comparative study of statistical models for yearly hurricane forecasting
}

\author{
  Pietro Colombo \\
  School of Mathematics and Statistics \\
  University of Glasgow, UK \\
  \texttt{Pietro.Colombo@glasgow.ac.uk} \\
  \And
  Raffaele Mattera\\
  Dipartimento di Scienze Sociali ed Economiche\\
  Universit\`a Roma la Sapienza, IT \\
  \texttt{raffaele.mattera@uniroma1.it} \\
  \And
  Philipp Otto \\
  School of Mathematics and Statistics \\
  University of Glasgow, UK \\
  \texttt{Philipp.Otto@glasgow.ac.uk} \\
}

\begin{document}
\maketitle

\begin{abstract}
In this paper, we study the problem of forecasting the next year's number of Atlantic hurricanes, which is relevant in many fields of applications such as land-use planning, hazard mitigation, reinsurance and long-term weather derivative market. Considering a set of well-known predictors, we compare the forecasting accuracy of both machine learning and simpler models, showing that the latter may be more adequate than the first. Quantile regression models, which are adopted for the first time for forecasting hurricane numbers, provide the best results. Moreover, we construct a new index showing good properties in anticipating the direction of the future number of hurricanes. We consider different evaluation metrics based on both magnitude forecasting errors and directional accuracy.

\end{abstract}


\section{Introduction} 
\label{sec1:introduction}
In this paper, we investigate the complex problem of forecasting the yearly number of hurricanes. For hurricanes, we intend a tropical cyclone in the Atlantic or East Pacific Oceans with winds exceeding 74 mph \citep[64 knots, 33 meters per second, see][]{world1993global}. Forecasting the frequency of hurricanes is crucial, especially because the average number of hurricanes occurring each year is increasing over time \citep{burn2015atlantic,sazcha2019increases}. Albeit some authors explained this evidence as part of a natural multi-decadal cycle \citep{chylek2008multidecadal,goldenberg2001recent}, many studies suggest that the increase in the yearly number of hurricanes is linked to climate change \citep[e.g. for a discussion see][]{mann2006atlantic,elsner2006evidence}. Questions remain about the factors that cause increased hurricane activities.  While the peak wind speed of a hurricane is the most well-known parameter used to describe the storm’s wind field, a full description of a hurricane is complex and involves possibly thousands of parameters ranging from ocean temperatures and salinities at various depths, atmospheric conditions such as temperatures, humidity, velocities, and so forth \citep{iman2006statistical,moharana2023recent,russell2020investigating}.

Monitoring and understanding the relevant predictors for hurricane activity has historically been challenging due to the physical systems' complexity. Tropical climate phenomena such as the El Niño–Southern Oscillation (ENSO) and the Atlantic Multidecadal Oscillation (AMO) contribute significantly to the variability in Atlantic hurricane occurrences \cite{klotzbach2017north}. Moreover, a stronger West African monsoon has been associated with more intense tropical storms and hurricanes due to factors such as stronger easterly waves, weaker vertical wind shear, and warmer sea surface temperatures \cite{bell2006leading}. Additionally, \cite{landsea1992strong} found a significant relationship between Gulf of Guinea rainfall during August–November and Atlantic hurricanes the following year. The Colorado State University (CSU) incorporates specific predictors in their seasonal outlooks to forecast Atlantic hurricane activity, including the low-level wind flow across the Caribbean Sea and sea surface temperature anomalies in the northeastern subtropical Atlantic. These predictors are chosen because they are closely linked to atmospheric and oceanic conditions that influence Atlantic hurricane activity, such as trade wind strength, pressure patterns, and the configuration of the African Easterly Jet (AEJ). The NOAA's seasonal hurricane outlooks for the North Atlantic basin rely on various climate factor predictors, including ENSO, AMO, and the Tropical Multidecadal Signal (TMS). The University College London in the United Kingdom has also been issuing public outlooks for seasonal tropical cyclone activity in the North Atlantic since December 1998. The advancements in yearly hurricane forecasting owe much to the accumulated experience of specialised centres and their dedication to improving forecast accuracy and reliability.

Historically, hurricane studies focused on forecasting their paths \citep{krishnamurti2016review}, even though the need for accurate forecasts of hurricane counts is relevant because it aids risk management in assessing potential damages to infrastructure and better mitigating human vulnerability. Good forecasts may enhance community preparedness and emergency response coordination, representing a useful tool for policymakers \citep{letson2007economic}. Moreover, accurate hurricane forecasting is also relevant to the business domain. Climate risk evaluation, in particular, is of paramount importance for investments, supply chain management and demand planning \cite{keenan2019climate}. For example, manufacturing companies rely on accurate forecasts to anticipate disruptions in the supply chain, which can be due to severe weather events. In the financial domain, accurate hurricane forecasts are essential for managing risk in derivatives investments tied to weather-related events. Weather derivatives contracts allow investors to hedge against financial losses resulting from adverse weather conditions, such as hurricanes \citep{meyer2014novel}. However, the effectiveness of these derivatives depends on the accuracy of weather forecasts. 
Initiatives like the Agora prediction market\footnote{For more details visit the CruciaLab website \url{https://www.crucialab.net/}}, which aims to aggregate outputs from quantitative models and expert tacit knowledge, represent promising avenues for leveraging insights from multiple dimensions and stakeholders. By generating high-resolution joint probability distributions, participants can make informed decisions and trade programmatically in response to changing climate dynamics. Simultaneously, the prices on the prediction market should reflect an accurate forecast of the hurricanes based on all traders' investment decisions. In this sense, the prices on the prediction are ensemble forecasts of (hidden) prediction models. A last noticeable example is represented by insurance and reinsurance firms. If hurricanes are underestimated, these companies may face unexpectedly high claims payouts, potentially amounting to billions of dollars in losses \citep{lyubchich2017can,philp2019issues}. Such losses can severely impact the financial stability of these firms, leading to potential disruptions in the insurance market and increased premiums for policyholders.


Different statistical approaches have been proposed to deal with the hurricane count forecasting problem. \cite{elsner1998multi} proposed the using a combination of different ARIMA models, while \citep{jagger2010consensus} demonstrates that the use of forecast combination of different GLM models with different predictors improves prediction accuracy in out-of-sample. \cite{daneshvaran2012atlantic} studied the performance of the ARIMA-X model, while \citep{daneshvaran2012atlantic,chang2018hurricane} more recently proposed to augment the ARIMA model with a predictor constructed with PCA. However, the use of Poisson distribution is the most natural choice for modelling the annual hurricane counts, and indeed, many authors followed this approach to describe hurricane counts \citep[e.g. see][]{parisi2000seasonality,elsner2008improving,xiao2015modeling,livsey2018multivariate}. Therefore, the INGARCH model represents a natural framework for modelling hurricane counts \citep{cui2016conditional}, even if some other authors \citep[e.g.][]{villarini2010modeling} have also considered models based on the negative binomial distribution to account for possible over-dispersion. In recent years, however, deep learning techniques have been used for hurricane count prediction tasks, particularly Recurrent Neural Networks (RNN) and Long Term Short Memory (LSTM) neural networks \citep{chen2020machine}.


In this paper, we introduce three main novelties. First, we present a forecasting experiment based on a probabilistic approach \citep[e.g. see][]{browell2020probcast} and, second, develop a novel index construction specifically tailored to assess Atlantic hurricane activity. Third, we systematically compare different statistical models and evaluate their efficacy across various performance metrics, including absolute accuracy, quantile accuracy, and directional accuracy. Directional accuracy, in particular, refers to the ability of a forecasting model to correctly predict the direction of change in the variable of interest \citep{granger2000economic}. It is widely adopted in economics \citep{blaskowitz2011economic}, but still overlooked in environmental sciences. Our interest in directional accuracy assessment is due to the evidence that statistical models based on past observations accurately predict the number of yearly occurrences of hurricanes only for certain years. Accounting for the right direction of change in the number of hurricanes is essential since it allows for the implementation of suitable adjustments. 

We advocate for the adoption of simpler modelling approaches, as they often guarantee greater interpretability and sensitivity, which are crucial for effective decision-making in the face of climate-related risks. A systematic approach was adopted to leverage past experiences and ensure reproducibility and transparency. This involved gathering data related to hurricane activity from specialised centres. 
Relevant predictors were then selected based on both statistical analyses \citep[e.g. stepwise approach, as in][]{knaff2007statistical} and observational evidence. With our approach, we try to leverage simplicity, with the need to provide different scenarios and uncertainty evaluations and accuracy.  We find that a simple quantile forecast might be beneficial for considering potential underestimation or overestimation in the decision-making process.

The rest of the paper is structured as follows. Section \ref{sec:data} shows the data associated with the paper, that is, the number of hurricane time series to be forecast and the relevant predictors used for this aim. In particular, Section \ref{sec:pp} discusses the rationale and the construction of the index used for improving forecast accuracy of the number of hurricanes occurring in the next year. Section \ref{sec:methods} provides a discussion about the methodology implemented, both the statistical methods used for forecasting and the details about the out-of-sample forecasting accuracy assessment. Section \ref{sec:results} shows the main results, while Section \ref{sec:final} concludes with final remarks and future research direction.

\section{Data}
\label{sec:data}

\subsection{Hurricane counts}

\begin{figure}[!htb]
    \centering
    \includegraphics[width=0.8 \textwidth]{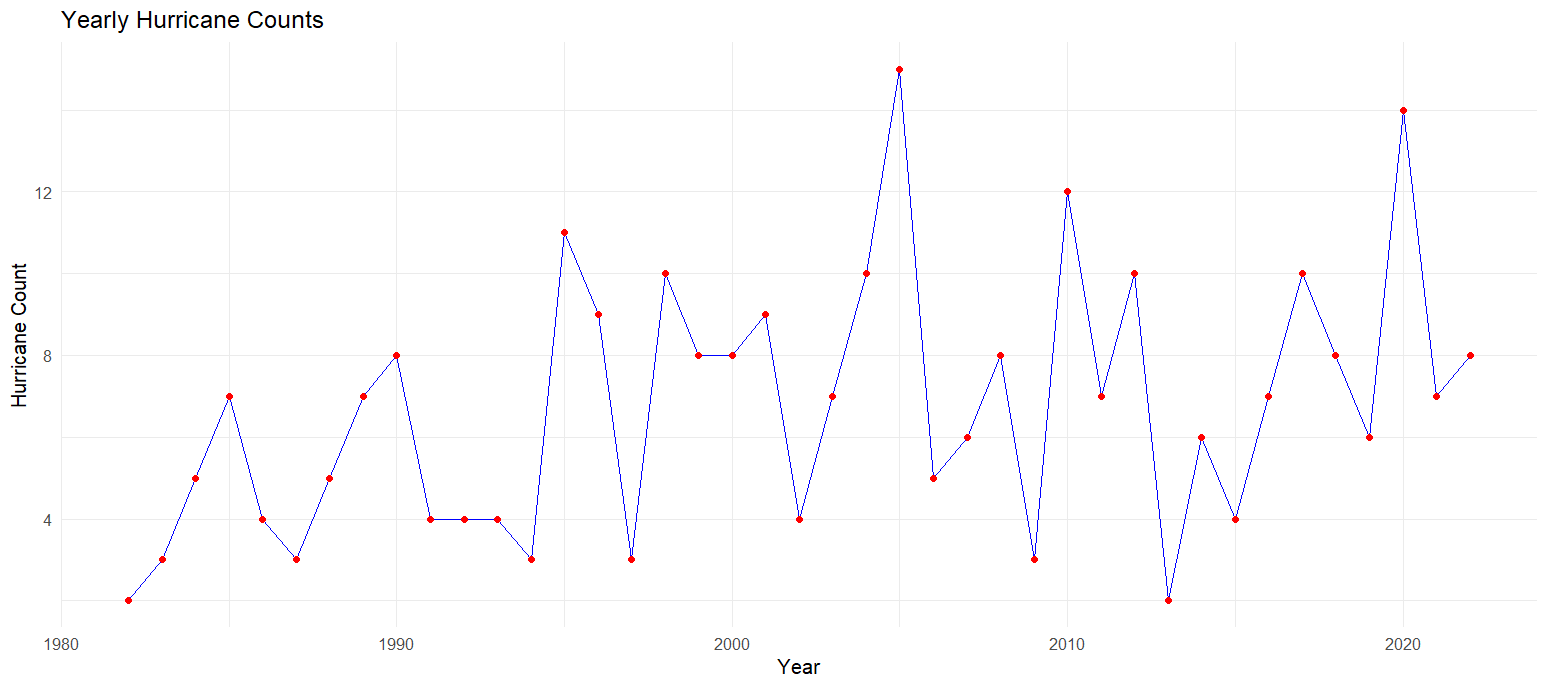}
    \caption{The figure depicts the yearly hurricane counts time series. On $x$-axis the year; while on the $y$-axis the yearly counts.}
    \label{fig:HurricaneCounts}
\end{figure}

Yearly hurricane counts are available for download from the Website at \cite{nhc2024}\footnote{We refer to the provided GitHub \url{https://github.com/Pietrostat193/Hurricane-forecasting} repository for sample R and Python code.}. The time series depicted in Figure \ref{fig:HurricaneCounts} spans from 1981 to 2022 and exhibits a modest upward trend interspersed with cyclical patterns every four to six years. This evidence is consistent with the previous studies highlighting the existence of cyclical patterns in the hurricanes' time series. Moreover, Figure \ref{fig:HurricaneCounts} shows a substantial increase in the variability of the number of hurricanes from the year 1994. Notably, certain years, such as 2005 and 2020, recorded an extraordinary number of hurricanes, exceeding 12 hurricanes per calendar year. In contrast, many other years typically saw fewer than four hurricanes annually. 

\noindent We are interested in forecasting next year's counts using statistical models and relevant predictors. The next sub-section describes the selected variables in the light of previous studies, while Section \ref{sec:pp} introduces a novel predictor.

\subsection{Hurricane counts predictors}\label{sec:predictors}

A systematic strategy was employed to use previous experiences and guarantee both reproducibility and transparency. This involved gathering data related to hurricane activity from specialised centres such as the National Hurricane Center (NHC) website\footnote{See \url{https://www.nhc.noaa.gov/}}, the Barcelona Supercomputing Center's seasonal hurricane predictions website \footnote{\url{https://seasonalhurricanepredictions.bsc.es/}}, Colorado State University's real-time hurricane data platform\footnote{\url{https://tropical.atmos.colostate.edu/Realtime/index.php?arch&loc=northatlantic}}, and the NOAA Earth System Research Laboratory's ENSO Dashboard \footnote{\url{https://psl.noaa.gov/enso/dashboard.html}}. We select an initial set of possibly relevant predictors based on previous studies.
As discussed in Section \ref{sec1:introduction}, tropical climate phenomena, particularly the El Niño-Southern Oscillation, play a significant role in forecasting hurricane activity. \cite{elsner2001secular} demonstrated a long-term relationship between ENSO patterns and the frequency of hurricanes.
Specifically, the ENSO is a climate phenomenon characterised by irregular fluctuations in sea surface temperatures (SSTs) and atmospheric pressure in the equatorial Pacific Ocean. It is one of the most significant drivers of climate variability on a global scale, affecting weather patterns, ocean conditions, and ecosystems around the world. The ENSO has two main phases:

\begin{enumerate}
    \item \textbf{El Niño}: This phase occurs when warmer-than-average sea surface temperatures develop in the central and eastern equatorial Pacific Ocean. It is typically associated with changes in atmospheric circulation patterns, such as the weakening of the trade winds and the shifting of the jet stream. El Niño events can lead to various impacts, including increased rainfall in some regions (e.g., South America and western North America), droughts in others (e.g., Australia and Indonesia), and disruptions to fisheries and agriculture.
\item 
\textbf{La Niña}: This phase is characterised by cooler-than-average sea surface temperatures in the central and eastern equatorial Pacific Ocean. La Niña events often bring opposite impacts to those of El Niño, such as enhanced rainfall in the western Pacific and drier conditions in parts of South America. The atmospheric circulation patterns during La Niña events tend to be stronger than normal, with stronger trade winds and a more active jet stream.
The transition between El Niño, La Niña, and neutral conditions (neither El Niño nor La Niña) is known as the ENSO cycle. This cycle typically occurs every 2 to 7 years but can vary in duration and intensity. ENSO events can have significant implications for various sectors, including agriculture, water resources management, energy production, and disaster risk reduction. Therefore, monitoring and understanding ENSO dynamics are crucial for worldwide climate prediction, risk management, and adaptation efforts. 
\end{enumerate}

\noindent Another fundamental predictor is the ESPI index, which is highly connected to the ENSO dynamic as shown in \cite{curtis2003evolution}. The index is derived from rainfall anomalies observed in two specific geographic regions: one situated in the eastern tropical Pacific (between 10°S and 10°N, and 160°E to 100°W), and the other located over the Maritime Continent (also between 10°S and 10°N, but spanning 90°E to 150°E). Initially, the process involves shifting a 10° by 50° block across each designated area. Subsequently, the minimum and maximum values within all possible blocks are determined for each region. These extremities are then amalgamated to calculate both an El Niño precipitation index (EI) and a La Niña precipitation index (LI). Finally, the EI and LI are integrated to establish the ESPI index.
Another index\footnote{Available here \url{https://psl.noaa.gov/enso/dashboard.html}} used in our model is the Zonal Average Temperature at specific atmospheric pressure level, specifically at 500 millibars, it refers to deviations from a long-term average. Table \ref{tab:index_descriptions} depicts the climate indices available at the Noaa Website (\cite{nhc_website}) that have been used as the starting point of our analysis. We performed a series of exploratory analyses looking for patterns between yearly hurricane occurrence and the annual averages of the climate indices. The exploratory analysis incorporated both data reduction techniques, such as PCA or multidimensional scaling, to uncover the association between a group of indices and bivariate plots. The latter was followed by a stepwise predictors selection, which ended up with the uncovering of two variables: the lagged 
standard deviation of La Niña,  and the lagged Zonal Average Temperature. The selection was particularly convenient since was based only on lagged information. Several attempts based on performance metrics have been made to include additional predictors but never managed to obtain performance better than those shown in table \ref{tab:evaluation_metrics}. Table\footnote{Note that the table is  copy of the dashboard of available at \cite{psl_enso_dashboard}.} \ref{tab:index_descriptions} provides the list of variables considered for the hurricane forecasting task.

\begin{table} 
    \centering
    \caption{Considered explanatory variables and descriptions}
    \small
    \begin{tabular}{ p{3cm} p{8cm} }
    \toprule
    \textbf{Name} & \textbf{Description} \\
    \midrule
    MEI V2 & A multi-variate index of ENSO using SST, winds, SLP, and OLR. Calculated using the JRA55 reanalysis dataset and the NCEI OLR. Produced at NOAA PSL. \\
    \hline
    Niño 4 & SST anomalies averaged over the NINO4 region 5°North-5°South; 160°East-150°West. Total SSTs also available for this region. Calculated from the Monthly NOAA ERSST V5 (at NOAA/CPC)\\
    \hline
    Niño 3.4 & SST anomalies averaged over the NINO34 region 5°North-5°South; 170-120°West. Total SSTs also available for this region. Correlates well with teleconnections to North America. Calculated from the Monthly NOAA ERSST V5 (at NOAA/CPC). \\
    \hline
    Niño 3 & SST anomalies averaged over the NINO3 region 5°North-5°South; 150°West-90°West. Total SSTs also available for this region. Calculated from the Monthly NOAA ERSST V5 (at NOAA/CPC).  \\
    \hline
    Niño 1.2 & SST anomalies averaged over the NINO1 and NINO2 regions 0-10°South; 90°West-80°West (eastern most of the Niña indices). Total SSTs also available for this region. Calculated from the Monthly ERSST V5 (at NOAA/CPC). \\
    \hline
    ONI & Oceanic Niño Index: 3-month moving average of ERSST.v4 SST anomalies in the Niño 3.4 region (5°N-5°S, 120°-170°W). Calculated from the ERSST V5 (at NOAA/CPC).  \\
    \hline
    BEST & Bivariate El Nino- Southern Oscillation Index: The Niño 3.4 SST and SOI are normalised and combined. SST is from the HadISST1.1. SOI is from NOAA/CPC. Produced at NOAA PSL. \\
    \hline
    SOI & Southern Oscillation Index: Difference between standardised Darwin and standardised Tahiti surface pressure values. Represents the atmospheric component of the ENSO. Sign is opposite that of the Niño indices and is noisier. From NOAA/CPC.  \\
    \hline
    TNI & Trans-Niño Index: Standardised Niño 1+2 minus the Niño 4 with a 5-month moving average applied (restandardised). Represents the gradient of the SST in the ENSO region of the tropical Pacific. Calculated from the Monthly HadISST1.1 dataset.  \\
    \hline
    PDO & Pacific Decadal Oscillation: leading principal component of monthly SST anomalies in the North Pacific Ocean, poleward of 20N (global SST mean removed). Calculated from the NOAA ERSSTV5, COBE SST, and HadISST1.1 (NOAA/PSL).  \\
    \hline
    PNA & Pacific North American Pattern. This version is calculated at NOAA/CPC. Based on EOF's calculated from monthly anomalies of 500mb height from the NCEP Reanalysis. \\
    \hline
    OLR & Outgoing Longwave Radiation (OLR) area averages over the central equatorial Pacific (160°E-160°W). OLR is a good measure of convection. Negative OLR represents increased convection. Calculated at NOAA/CPC.  \\
    Heat Content & Tropical Pacific integrated temperature anomalies (0-300m) 160°E–80°W. Calculated at NOAA/CPC from the GODAS dataset.  \\[.4cm]
    \hline
    200mb Zonal Winds & 200mb Zonal Wind anomalies 2.5S-2.5N; 165W-110W. Calculated at NOAA/PSL from the NCEP R1. \\
    \hline
    ESPI Precip Index & ENSO Precipitation Index. The index is based on rainfall anomalies in two rectangular areas, one in the eastern tropical Pacific (10°S-10°N, 160°E-100°W) and the other over the Maritime Continent (10°S-10°N, 90°E-150°E). The first step of the procedure involves moving a 10° by 50° block around each box; the minimum and maximum values of all possible blocks are obtained for each box, and these are combined to estimate an El Niño precipitation index (EI) and a La Niña precipitation index (LI). The EI and LI are, in turn, combined to create the ESPI index. Finally, the ESPI index is normalised to have zero mean and unit standard deviation. Calculated at UMD.  \\
    \bottomrule
    \end{tabular}
    \label{tab:index_descriptions}
\end{table}


\begin{figure}[!htb]
    \centering
    \includegraphics{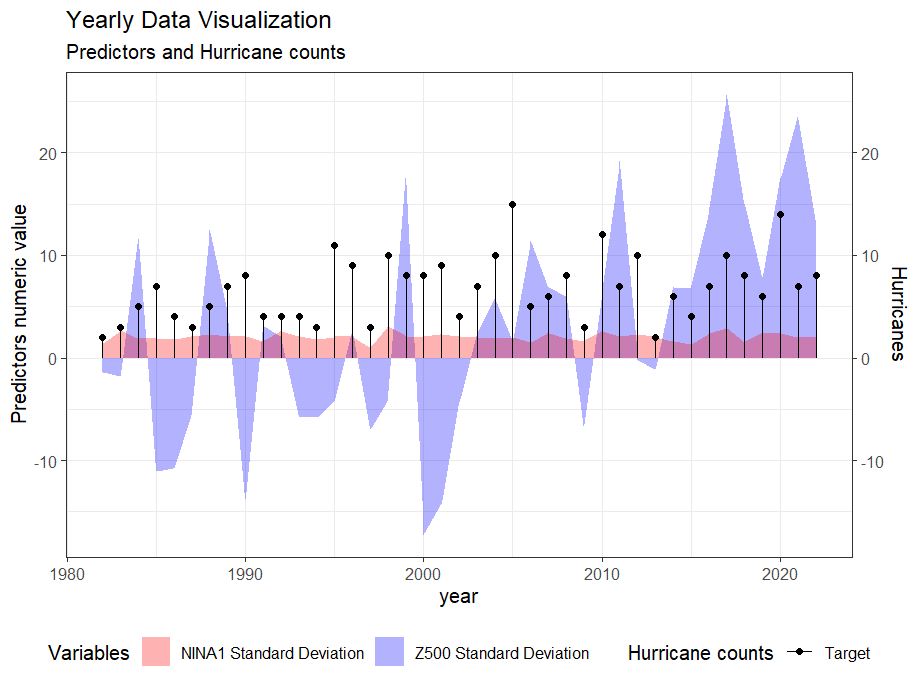}
    \caption{The figure shows the two selected predictors overlay to the hurricane counts with a stepwise approach. On $x$-axis the calendar year, while on the $y$-axis the numeric values of the selected indices and the hurricane counts.}
    \label{fig:Predictors and Hurricane counts}
\end{figure}

\subsection{A novel predictor: a Co-Movement Index}
\label{sec:pp}
\begin{figure} 
    \centering
    \includegraphics[width=0.8\textwidth]{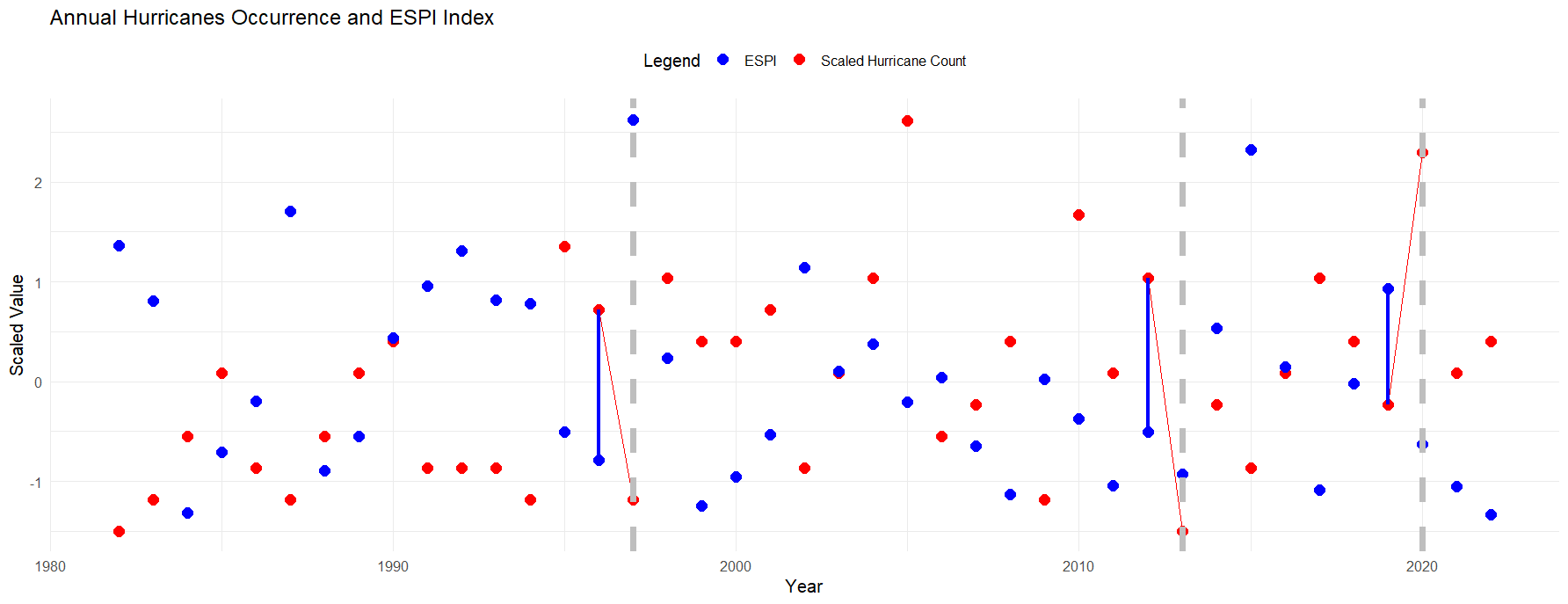}
    \caption{This figure compares the scaled occurrences of hurricanes with the ESPI index. Vertical grey lines denote ``anomalous years'', emphasizing deviations from typical patterns. An oblique red segment highlights year-over-year changes, while vertical blue bars illustrate scale discrepancies between the two variables.}
    \label{fig:HurVSEsp}
\end{figure}

Upon seeking further information to enhance our model's performance, we discovered that the models were only underperforming in specific years. These years exhibited anomalous counts compared to prior observations. A simple yet convenient intuition is that incorporating a predictor that accounts for the anomalous years would have improved the model prediction accuracy. The latter posed the basis for the creation of the Pseudo-Predictor (PP). More specifically, it was observed that the ESPI index appears to be a decent predictor of hurricane counts over the years, as shown in Figure \ref{fig:HurVSEsp}, with a negative correlation of 0.48$\%$. However, significant disparities exist between the two datasets, especially during abnormal years when hurricane counts sharply differ from the previous year. For example, in 1997, 2013, and 2020, noticeable deviations occurred compared to the past (highlighted by red diagonal lines in Figure \ref{fig:HurVSEsp}). Interestingly, during these years, the difference between the ESPI index and the previous year's hurricane counts (blue segments in Figure \ref{fig:HurVSEsp}) also showed considerable differences. Therefore, leveraging this lagged discrepancy could be helpful in predicting the anomalous years and enhancing the overall prediction accuracy.

To define an additional so-called PP, we continuously estimate a simple linear regression model between the number of yearly hurricanes of the last 29 years and the temporally lagged discrepancy between scale version ESPI and the number of hurricanes (NH) of the previous year. In other words, the linear regression model is estimated using a rolling-window approach to predict the number of hurricanes this year, which will be used in the prediction models for the following year. Let the estimated coefficients of this linear regression model be $\hat{\beta}_{0,t}$ and $\hat{\beta}_{1,t}$. Then, the PP is defined as 

\begin{equation}
    \text{PP}_t =  \hat{\beta}_{0,t} + \hat{\beta}_{1,t} \text{DIS}_{t-1} \, . 
\end{equation}
\\
where $\text{DIS}_{t}=\widetilde{\text{ESPI}}_{t}-\widetilde{\text{NH}}_{t}$, given that $\widetilde{\text{ESPI}}_t$ is the standardised value of the ESPI index at the year $t$ and $\widetilde{\text{NH}}_t$ is the standardised hurricanes counts at the year $t$.
The PP index displays an 86\% negative correlation with the annual occurrence of hurricanes, a figure significantly higher than that of any other climatic indices. Additionally, the PP index is constructed using lagged information, making it well-suited for forecasting purposes.

\section{Methodology}
\label{sec:methods}
In this section, we present the experimental study, illustrating the models we aim to test to predict yearly hurricane occurrence, the metrics, and the experimental design.

\subsection{Models}

In the following section, we provide a comprehensive overview of all considered forecasting models. We selected models based on several criteria:

\begin{enumerate}
\item \textbf{Standard time series models}: Since the problem at hand primarily revolves around the prediction of yearly hurricane occurrences, we recognised the importance of incorporating standard time-series models. These models are well-suited for addressing the inherent temporal dependencies present in the data. Initially, we will consider simple real-valued time series models.

\item \textbf{Integer-valued time series models}: Given that yearly occurrences of hurricanes are an integer-valued time series--with some years having particularly low counts--we deemed it essential to include count models in our selection. This allows us to assess the appropriateness of using count models for forecasting tasks. Compared to the above-mentioned real-valued time series, which forecast the expected number of hurricanes as a real-valued quantity that needs to be rounded to integers to get an appropriate forecast, integer-valued time series models emulate an integer forecast directly (based on the modelled real-valued expectation).
\item \textbf{Probabilistic forecast models}: We aimed to incorporate probabilistic forecast models to provide distributional predictions and allow for asymmetry and more likely extremes. The other models of the previous categories can provide distributional forecasts, which are solely based on the assumed error distribution. By contrast, for quantile regression approaches, the distribution quantiles of interest are separately modelled. In this context, it is crucial not only to understand the shape of the distributions but also to assess their informativeness. This involves considering whether the distributions adequately account for potential underestimation or overestimation of hurricane occurrences, thus providing a more comprehensive perspective on the associated risks. This could be of particular importance for prediction markets, such as the Agora market.
\item \textbf{Machine-learning approaches}: We considered the inclusion of so-called ``black-box models'' due to their prevalent use in the literature for providing forecasts. As discussed in the introduction (Section \ref{sec1:introduction}), black box technologies are often employed to achieve high prediction accuracy. However, they may lack the necessary scientific transparency required for robust risk evaluation. Thus, there exists a trade-off with black box technologies. While they may deliver accurate predictions, they might also lack the explainability necessary for effective risk assessment. We specifically chose artificial neural networks with a Long Short-Term Memory (LSTM) layer as a representative black box technology, which also accounts for the inherent temporal dependence in the data.
\end{enumerate}

To ensure a fair comparison of the predictive capabilities across all models, we selected identical sets of regressors for each, thereby ensuring that variations in performance stemmed solely from the models' structures, not from differing input information. We used two specific sets of regressors: 
\begin{enumerate} 
    \item Standard baseline set: Standard deviation of yearly La Niña index at year $t-1$ and Zonal Temperature 500 anomalies at year $t-1$,
    \item Enhanced set: Standard baseline set and the proposed pseudo predictor $\text{PP}_t$.
\end{enumerate}
These choices demonstrated reasonable performance across different model types and alternative sets of regressors (see Table \ref{tab:index_descriptions} for an overview of all considered regressors). By opting not to perform model-specific selection, we prioritised uniformity over potentially achieving optimal performance for individual models. This approach highlights each model framework's inherent strengths and weaknesses, facilitating a clearer evaluation of their relative effectiveness and allowing us to draw more direct comparisons about model construction and its impact on forecasting accuracy.

Let $\{Y_t \in \mathbb{Z} : t \in \mathbb{Z} \}$ be the discrete time series of the yearly counts of Hurricanes across the Atlantic Ocean and $\{y_t \in \mathbb{Z} : t = 1, \ldots, T, \ldots, F \}$. be the observed series, where $T$ denotes the number of time points used for estimation of the models and the time points $t = T+1, \ldots, F$ are used for validation of the prediction performance. Moreover, let $\boldsymbol{X}_t = (x_{it})_{i = 1, \ldots, p}$ be the $p$-dimensional vector of exogenous regressors at time $t$ from the two covariate sets.

As a baseline time-series model, we considered a regression model with autoregressive integrated moving average (ARIMA) residuals, a pivotal tool in time-series forecasting. An ARIMA model assumes that the differenced series $(1-B)^d y_t$ with $d \in \mathbb{Z}_{0}^{+}$ and $B$ being the backshift operator (i.e., $B y_t = y_{t-1}$) is a stationary autoregressive moving average (ARMA) process. For instance, if $d = 1$, the first differences $\{y_t - y_{t-1}\}$ follow a stationary ARMA process. Thereby, the model allows us to capture both trends and cyclicality in the historical hurricane counts. More precisely, an ARIMA$(p, d, q)$-X model can be expressed as follows:

\begin{equation}
    \phi(B)(1-B)^d y_t = \theta(B) \epsilon_t + \boldsymbol{X}'_t \boldsymbol{\beta},
\end{equation}
\\
where $\phi(B) = 1 - \phi_1 B - \ldots - \phi_p B^p$ represents the autoregressive polynomial, $\theta(B) = 1 + \theta_1 B + \ldots + \theta_q B^q$ denotes the moving average polynomial, and $\{\epsilon_t\}$ is white noise process. This framework not only facilitates the understanding of past patterns but also enhances forecast accuracy by integrating past values (autoregressive part) and error terms (moving average part) to predict future outcomes. To choose the model orders/parameters $p$, $d$, and $q$, we employed the Hyndman-Khandakar algorithm \citep{hyndman2008automatic}. It determines the optimal number of differences ($0 \leq d \leq 2$) using the Kwiatkowski–Phillips–Schmidt–Shin (KPSS) unit root tests to find a stationary process. Then, the orders $p$ and $q$  are selected by minimizing the corrected Akaike information criterion, AICc, in a stepwise search procedure. Further, the model parameters are estimated by the quasi-maximum-likelihood (QML) approach.

Building upon our baseline model, we incorporated the integer-valued GARCH models with exogenous regressors (INGARCH-X) to analyse the count data of yearly hurricane occurrences. The INGARCH($p$, $q$)-X model treats the counts $Y_t$ as being conditionally Poisson distributed (i.e., conditioned on the past observations $Y_{t-1}, Y_{t-2}, \ldots$), with the intensity/mean parameter $M_t$ given by

\begin{equation}
E(Y_t | Y_{t-1}, Y_{t-2}, \ldots) = M_t = \omega + \sum_{i=1}^{p}\alpha_i Y_{t-i} + \sum_{j=1}^{q}\gamma_j M_{t-j} +  \boldsymbol{X}'_t \boldsymbol{\beta}    
\end{equation}
\\
with $\omega > 0$ and $\alpha_1, ...,\alpha_p,\gamma_1,...,\gamma_q \geq 0$. This formulation highlights the autoregressive nature of the counts and the direct impact of external factors. It effectively captures the autocorrelation inherent in the count data while incorporating the influence of exogenous variables that affect the occurrence of events. The conditional Poisson assumption allows for the explicit modelling of count variability based on past counts and external influences, as the variance is given by $M_t$ (however, note that this also implies equidispersion). For a more detailed overview of INGARCH models, we refer the interested reader to \citep[][Chapter 4]{weiss2018introduction}.

In addition to the time-series models for forecasting hurricane occurrences, we also include quantile regression (QR) models and quantile gradient-boosted regression trees  (QGBRT) in our comparative study for probabilistic forecasts. Quantile regression, unlike traditional regression methods that model the conditional mean of the dependent variable, focuses on estimating the conditional quantiles. That is, the conditional $\tau$-th quantile of the response variable is given by

\begin{equation}
 Q_{Y_t}(\tau | \boldsymbol{X}_t) = \boldsymbol{X}'_t \boldsymbol{\beta}(\tau),   
\end{equation}
\\
where $\boldsymbol{\beta}(\tau)$ represents the vector of quantile-specific coefficients. The parameters $\boldsymbol{\beta}(\tau)$ can be estimated by minimising 

\begin{equation}
 \sum_{t=1}^{T}  \rho_\tau(Y_t - \boldsymbol{X}'_t \boldsymbol{\beta}(\tau)),   
\end{equation}
\\
with $\rho_\tau$ being a loss function $\rho_\tau(x) = x (\tau - \mathbb{I}_{(-\infty, 0)}(x))$. Here, $\mathbb{I}_A(x)$ denotes an indicator function on the set $A$. For a general review/overview of quantile regression models, we recommend reading \cite{koenker2017quantile,cade2003gentle}.

In addition to these statistical models, we additionally considered methods from statistical/machine learning, starting with quantile gradient-boosted regression trees. This is an extension of gradient boosting methods that adaptively build ensemble models focussing on quantile loss functions $\rho_\tau(x)$. Instead of the linear influence of the predictors in the model above, this also includes nonlinearities and interactions among the predictors. That is, the model iteratively fits regression trees to minimise the loss function derived from the difference between the observed quantiles and the predicted quantiles

\begin{equation}
L(Y_t, \hat{Q}_{Y_t}(\tau | \boldsymbol{X}_t )) = \sum_{t=1}^T \rho_\tau(Y_t - \hat{Q}_{Y_t}(\tau | \boldsymbol{X}_t )),    
\end{equation}
\\
where $\rho_\tau(x)$ is a quantile loss function, which weights the residuals differently depending on whether they are above or below the predicted quantile, thus focusing the learning on different quantiles of the distribution of $Y_t$.  QGBRT have already been successfully applied in other environmental studies, see, e.g., \cite{shaziayani2021coupling,vasseur2021comparing}.

Incorporating deep learning capabilities, we finally utilise recurrent neural networks (RNNs) with Long Short-Term Memory (LSTM) layers to predict hurricane counts. RNNs with LSTM architecture are particularly suited for modelling time series, where the dependence of current data points on previous ones should be employed for forecasting. In the hurricane context, such approaches have been employed, for instance, for predicting the evolution and trajectories of hurricanes \citep{bose2022real,qin2021trajectory}.

Generally, the LSTM model for predicting hurricane counts $Y_t$ at time point $t$ (i.e., the final layer) can be defined as 

\begin{equation}
Y_t = f(\mathbf{W} \cdot [\boldsymbol{h}_{t-1}, \boldsymbol{X}_t ] + \boldsymbol{b}),    
\end{equation}
\\
where $\boldsymbol{h}_{t-1}$ is the hidden state from the previous time step (i.e., information from past time points), $\mathbf{W}$ and $\boldsymbol{b}$ are the weights and bias of the neural network, respectively, and $f$ is a (non-)linear activation function (e.g., linear, ReLU or softplus activation function).

The LSTM layer updates through several gates to regulate the flow of information, maintaining a memory cell $c_t$ that adjusts over time with so-called input, forget, and output gates. More precisely, the gates $i_t$, $f_t$, and $o_t$ are defined as 
\[
\begin{aligned}
i_t &= \sigma(\mathbf{W}_i \cdot [h_{t-1}, \boldsymbol{X}_t] + \boldsymbol{b}_i), \\
f_t &= \sigma(\mathbf{W}_f \cdot [h_{t-1}, \boldsymbol{X}_t] + \boldsymbol{b}_f), \\
o_t &= \sigma(\mathbf{W}_o \cdot [h_{t-1}, \boldsymbol{X}_t] + \boldsymbol{b}_o), \\
\tilde{c}_t &= \tanh(W_c \cdot [h_{t-1}, \boldsymbol{X}_t] + b_c), \\
c_t &= f_t \odot c_{t-1} + i_t \odot \tilde{c}_t, \\
h_t &= o_t \odot \tanh(c_t)
\end{aligned}
\]
with \(\sigma\) and \(\tanh\) representing the sigmoid and hyperbolic tangent functions, and \(\odot\) denotes element-wise multiplication.
In simple words, these gates decide how much information from the data to let into the cell, how much of the existing cell state to keep, and how much of the cell state to output to the hidden state $h_t$. In this way, the RNN can effectively learn dependencies over longer time horizons.

\subsection{Forecasting experiment and accuracy evaluation}

The experimental design aims to compare the effectiveness of different models in forecasting the next year's number of hurricanes, considering various predictor sets and evaluation metrics. We aim to evaluate if we can improve the accuracy and reliability of yearly hurricane activity predictions by the use of relatively simple approaches. For the out-of-sample evaluation, we split the sample, leaving the last eleven years of data for the testing. We evaluate the accuracy of one-step ahead forecasts, considering an expanding window approach. Therefore, given an initial window equal to the training set, we estimate the parameters of each statistical model and obtain the forecasts for the next year. Then, we move one year ahead and expand the estimation window. Using the new data, we re-estimate the models' parameters and obtain a new one-step ahead forecast. This approach is applied recursively until no new observations are available. We compare four model categories: simple time series (ARIMA), count time series (INGARCH), probabilistic approaches (QR and QGBRT), and machine learning (LSTM). For each considered model, we consider the predictors selected with the stepwise approach, as explained in Section \ref{sec:predictors}. Then, we evaluate the improvements in the forecasting performance that can be obtained considering the Pseudo Predictor (PP) developed in Section \ref{sec:pp}. In the case of LSTM, we adopt an ensemble approach of different trained models to both improve the out-of-sample performance and reduce the sensitivity to the starting values. Indeed, as shown by many previous studies, better forecasts in out-of-sample may be achieved by considering a combination of different neural network (NN) models instead of selecting a single NN that may be susceptible to poor initial values \citep{kourentzes2014neural,taieb2012review}. We apply the same reasoning to the LSTM and use the mean ensemble operator of 100 trained LSTM models, as the mean often outperforms more complex methods of combining forecasts \citep{naftaly1997optimal}.

We then test the accuracy of the statistical models in three main dimensions. The first is directional accuracy, which allows us to understand if hurricanes are increasing from one year to another. The directional accuracy (\textbf{DA}) metric measures the proportion of correctly predicted directional changes in a forecast compared to the actual values. Let us denote the index with $i$, referring to the $i$-th point in time in the testing set. The directional accuracy is defined as
\begin{equation}
    \textbf{DA} = \frac{{\sum_{i=2}^{n} \mathbb{I}_{\text{sign}(y_{i} - y_{i-1})} \left( \text{sign}(\hat{y}_{i} - \hat{y}_{i-1}) \right)}}{{n - 1}},
\end{equation}
where $\hat{y}_{i}$ is the forecasted value at time $i$, $y_i$ is the actual value at time $i$, $n$ is the total number of observations in the validation window $i = T+1, \ldots, F$ and $\mathbb{I}_A(x)$ is an Indicator function on a set $A$. In this equation, the numerator counts the number of instances where the sign of the difference between consecutive forecasted values (\(\hat{y}_{i} - \hat{y}_{i-1}\)) matches the sign of the difference between consecutive actual values (\(y_{i} - y_{i-1}\)). This count is divided by \(n - 1\) to normalise the directional accuracy metric to the range [0, 1], where 1 represents perfect directional accuracy.

The second metric is the Mean Absolute Error \textbf{MAE}, as we want to know how close we are to the actual yearly hurricanes. The Mean Absolute Error (\textbf{MAE}) is a commonly used metric to evaluate the accuracy of a forecasting model. It measures the average magnitude of errors between the predicted values and the actual values, that is,
\begin{equation}
    \textbf{MAE} = \frac{1}{n} \sum_{i=1}^{n} | \hat{y}_{i} - y_{i} |.
\end{equation}
In this equation, for each observation \(i\), the absolute difference between the forecasted value \(\hat{y}_{i}\) and the observed value \(y_{i}\) is calculated. These absolute differences are then averaged across all observations to obtain the MAE. The MAE provides a straightforward measure of the average magnitude of errors, where smaller values indicate better accuracy of the forecasting model. 

Finally, to evaluate the quality of the probabilistic forecast, we used the average pinball loss function (\textbf{APLF}). The average pinball loss function for each quantile at a given time $t$ represents the average loss incurred by a quantile forecast over a period of time $t$.  The pinball loss function is commonly used in quantile regression and forecast evaluation. It measures the deviation between the actual observed value $y$ and the forecasted quantile value $q$ at a specific quantile level $\tau$. The loss function is defined as
\begin{equation}
L(y_i, q_{\tau}) = 
\begin{cases} 
    (1 - \tau)(y_i - q_{\tau}) & \text{if } y_i \geq q_{\tau} \\
    \tau(q_{\tau} - y_i) & \text{if } y_i < q_{\tau} \, ,
\end{cases}
\end{equation}
where $y_i$ is the observed value, $q_{\tau}$ is the forecasted quantile value at quantile level $\tau$, and $\tau$ is the quantile level (e.g., $0.1$ for the 10th percentile or $0.5$ for the median). The average pinball loss function for each quantile at a given time $i$ is then calculated by taking the average of the pinball loss function values across all observations at that time $i$. This provides an overall assessment of the forecast accuracy across different quantiles at that specific time point. In summary, the average pinball loss function for each quantile at a given time helps to evaluate the performance of quantile forecasts by measuring the average deviation of forecasted quantiles from the observed values over time. The last metric is also applied to time-series and count models, where quantiles are calculated based on a normal distribution centred on the forecast and using the model's estimated standard deviation. To our best knowledge, this is a novelty, but our intuition is that such a metric would assess the suitability of normal uncertainty estimations as opposed to a more flexible quantile-based approach.

\section{Results}
\label{sec:results}

\begin{figure}
    \centering
    \includegraphics[width=0.8\textwidth]{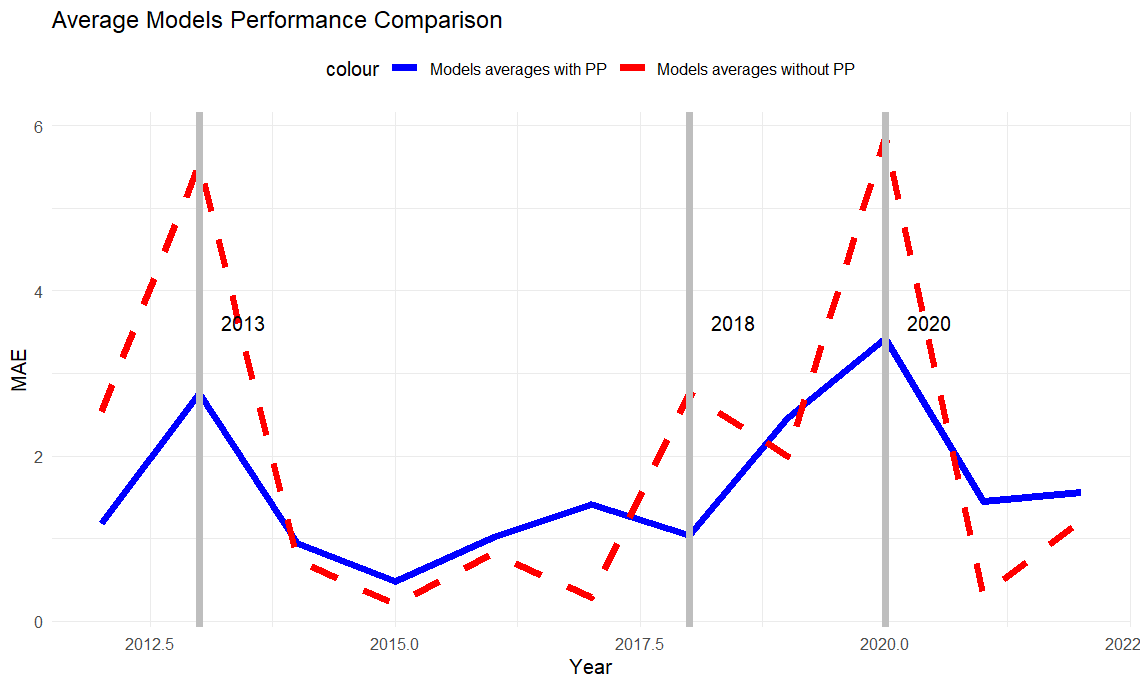}
    \caption{MAE test results for two sets of models. The first set encompasses the MAE averages of models incorporating the PP predictor (blue line), while the second set (dotted red line) represents the MAE averages of models excluding the PP predictor. Vertical lines emphasise anomalous years marked by substantial prediction errors across all models except those utilising the PP predictor.}
    \label{fig:Hurricane_predictionAVG}
\end{figure}

\begin{figure}
  \centering
\subfigure[Average PBLF for the ARIMA-PP model]{
    \includegraphics[width=0.8\textwidth]{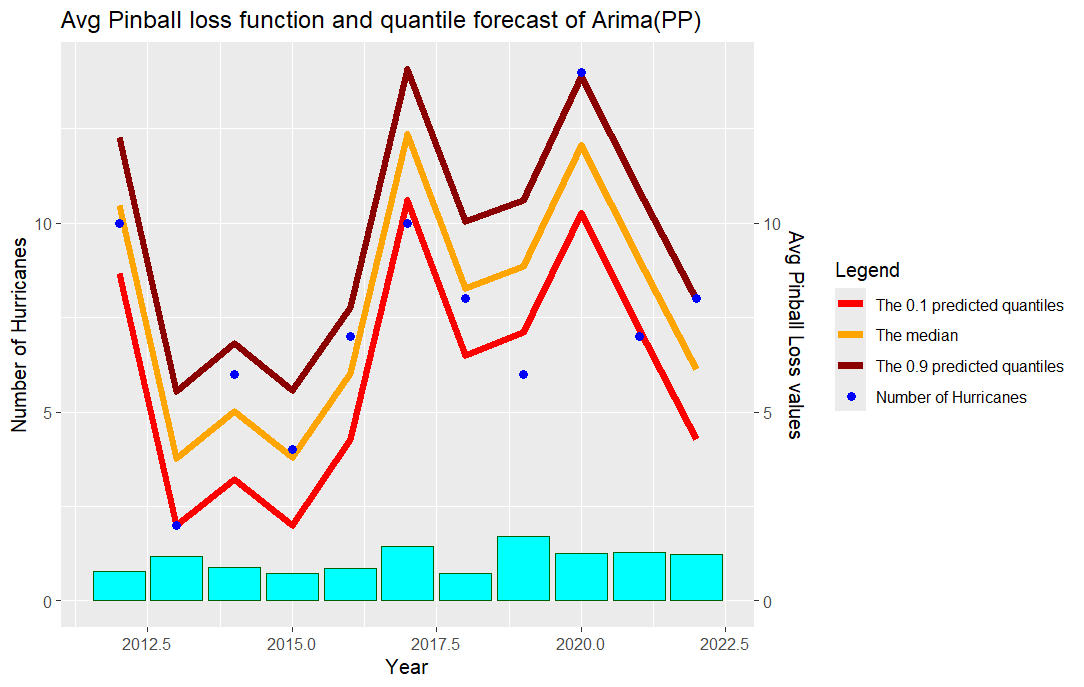}\label{fig:plot3}
}
\subfigure[Average PBLF for the QR-PP model]{
    \includegraphics[width=0.8\textwidth]{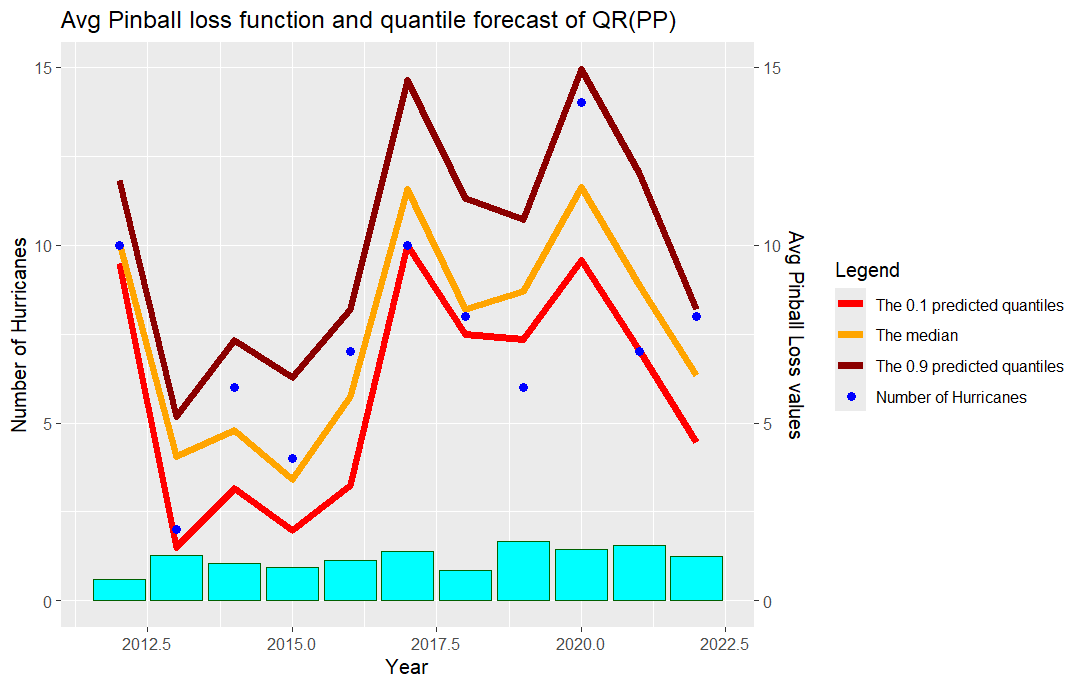}
    \label{fig:plot4}
}
  \caption{Yearly  0.1, 0.5, and 0.9 quantiles predictions for the yearly hurricane counts for the ARIMA+PP(top panel) and the QR+PP(bottom panel). The light blue histograms represent the yearly APLF. Note that in 2020 and 2013, the predicted quantiles of the QR regression correctly spotted the overestimation (2013) or underestimation (2020).}
  \label{fig:multipanelPrediction}
\end{figure}

Table \ref{tab:evaluation_metrics} offers a detailed look at different modelling methods and how well they perform, using various measures to gauge their effectiveness. We start our discussion with all models without the pseudo predictor. In this first setting, the Quantile Regression (QR) approach provides the best accuracy, with an MAE of 1.96, while the INGARCH shows the second-best performance with an MAE equal to 1.99. Therefore, a probabilistic approach provides more accurate forecasts in terms of point magnitude errors while it also provides information about other quantiles. Interestingly, the LSTM provides larger forecasting errors compared to much simpler models, suggesting that complex models do not ensure better predictions in out-of-sample. All the simpler models (ARIMA, INGARCH, QR), assume some forms of linear relationships. Even though this type of linear relationship may not fully account for the year-to-year variability in hurricanes, it is sufficient to deliver competitive results. In other words, when using non-linear relations for forecasting into the future (i.e., out-of-sample not included in the training domain), the added complexity does not necessarily translate into improved accuracy, and the models may fail to generalise well to unseen data. By contrast, linear relations constrain the forecasting discrepancies. This observation underscores the importance of model simplicity and robustness, especially when dealing with datasets where linear models have already proven effective. Thus, simpler models can provide reliable and competitive performance for forecasting tasks. That is, models like QGBRT and LSTM are designed to uncover complex relationships, which can be problematic if such relationships do not exist. For instance, in QGBRT, each successive tree is trained to correct the residuals of the previous ones, which increases the risk of the model interpreting noise as a signal, especially if the training dataset provides little informative value. Similarly, LSTM might also capture noise rather than substantive data. Generally, both models involve complex parameter tuning that requires specific studies; for example, selecting the optimal number of trees, tree depth, learning rate, and other hyperparameters is critical in boosting algorithms. Misjudging these parameters can lead to either underfitting or overfitting. Moreover, the interaction between these parameters can complicate the tuning process, making it more challenging to optimise the models effectively. Therefore, machine learning tools may provide better results in out-of-sample if higher frequency data (e.g. monthly hurricanes) are considered for the forecasting experiment. 

When considering Directional Accuracy (DA), the ARIMA performs worse compared to other methods, indicating that it fails to correctly predict the direction of change in hurricane counts. Combined with the associated large MAE loss, the DA indicates that the ARIMA model provides very poor forecasts of the number of counts while missing the direction. The best model in terms of directional accuracy belongs to the probabilistic class, that is, the QR, which has a DA measure of 0.7. Both MAE and DA are based on point forecasts, but in this paper, we also compare the forecasting performance in a distributional sense. Comparing APLF among these models helps assess the benefit of using normal quantiles (ARIMA, INGARCH) versus more flexible ones (QR, QGBRT). While ARIMA and INGARCH show lower APLF, indicating normal quantiles are not inappropriate, they lack information on potential model underestimation/overestimation. We excluded LSTM  from this analysis as they were not designed for uncertainty prediction/quantification. The QR models, on the other hand, provide insightful quantiles. For instance, in Figure \ref{fig:multipanelPrediction}, extreme quantiles like 0.1 in 2013 and 0.9 in 2020 correctly signal overestimation and underestimation, respectively.  On the other side, ARIMA model quantiles have equal distances, which might be good overall but lack information on potential under/overestimation. 

We notice significant improvements in forecasting accuracy for all the methods, particularly for the probabilistic approaches. For example, the MAE for QR decreases from 1.96 to 1.41 when PP is applied, indicating a substantial reduction in magnitude errors. Similarly, the QGBRT model sees a decrease in MAE from 3.25 to 1.58 with PP. This demonstrates the effectiveness of PP in enhancing the accuracy of probabilistic forecasts. Furthermore, probabilistic approaches with PP exhibit superior performance in terms of probabilistic forecast metrics like APLF. Among the time-series models, ARIMA+PP stands out for its very low MAE and APLF. This result is surprising since the ARIMA model without PP was among the models performing worst. Similar results are obtained with INGARCH+PP. 
Figure \ref{fig:Hurricane_predictionAVG} illustrates the average MAE of the top-performing models (ARIMA, QR, INGARCH). The comparison is shown between scenarios where the PP predictor is included (blue line) and where it is omitted (red dotted line).The image shows that the benefit of the PP predictor is particularly important in specific years, such as the years 2013 and 2020 that are notably anomalous. Moreover, for the year 2018, it follows a significant discrepancy between the ESPI and the yearly standardised hurricane count, which suggests the predictive usefulness of the PP predictor. Indeed, while it is true that anomalous years often precede a large $\text{DIS}{t}$ variable, the converse is not always true; not all non-anomalous years are preceded by low values of the $\text{DIS}{t}$ variable. However, incorporating this predictor in other years may even reduce the model performance.


\begin{table} 
\centering
\caption{Evaluation metrics for all considered models}
\label{tab:evaluation_metrics}
\begin{tabular}{llccc}
\toprule
\textbf{Model category} & \textbf{Model} & \textbf{MAE} & \textbf{DA} & \textbf{APLF} \\
\midrule
Time-series           &ARIMA+PP     &1.42  &0.8   & 1.09\\
                      &ARIMA        &2.15  &0.6   & 2.09 \\
                      \hline
Count Model           &INGARCH+PP   &1.48  &0.8   & 1.12\\
                      &INGARCH      &1.99  &0.7   & 1.97 \\
                      \hline
Probabilistic forecast&QR+PP        &1.41  &0.8   & 1.19\\
                      &QR           &1.96  &0.7   & 1.7 \\
                      &QGBRT+PP     &1.58  &0.7   & 1.22  \\
                      &QGBRT        &3.25  &0.2   &  1.82\\
                      \hline
Black Box Machine learning&LSTM+PP  &2.1   & 0.8  & NA  \\
                          &LSTM     &2.26   & 0.8   & NA    \\
\bottomrule
\end{tabular}
\end{table}

\begin{figure} 
    \centering \includegraphics[width=0.8 \textwidth]{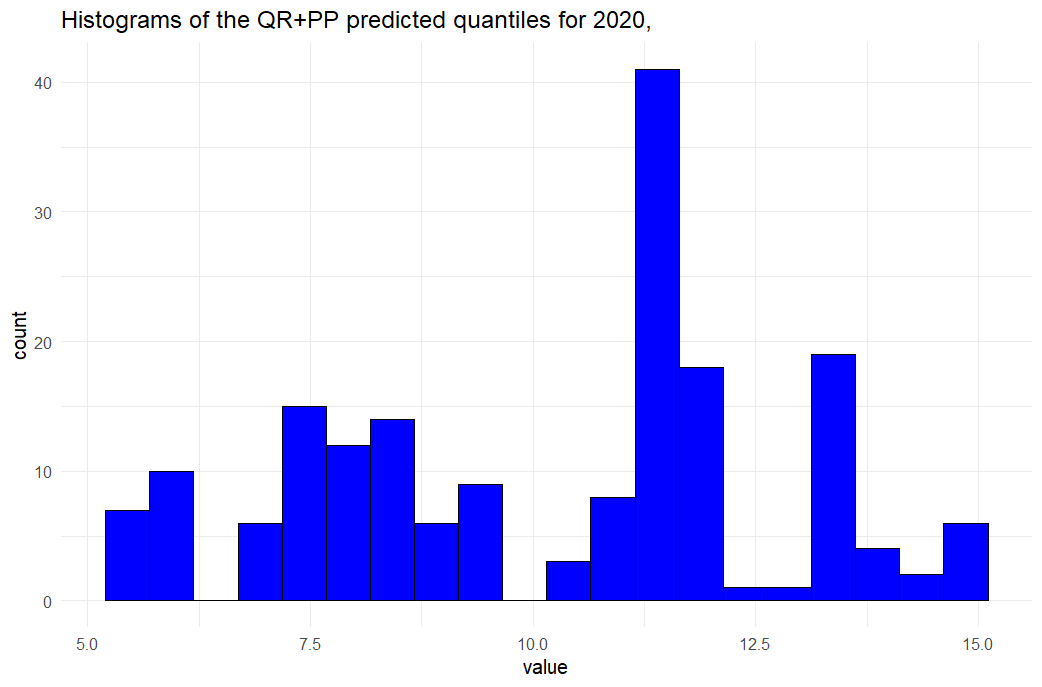}
    \caption{Predicted quantiles distribution of QR+PP for 2020. Note that high weight is given on the right tail since the model correctly spotted the underestimation.}
    \label{fig:quantiles_dis}
\end{figure}

To further evaluate the calibration and accuracy of these models, we turn to the advantage of using a predictive series of quantiles for each time point. This approach allows us to assess the calibration of the model using the ``The Probability Integral Transform'' (PIT) value, which represents the cumulative probability of the observed value under the forecasted probability distribution. To be precise, we can distinguish the following cases: 
\begin{itemize}
    \item \textbf{PIT = 0.5:} This means the observed value is exactly at the median of the forecast distribution. The forecast is perfectly centred around the observed value, suggesting good calibration for this particular instance.
    \item \textbf{PIT $<$ 0.5:} If the PIT value is less than 0.5, the observed value falls in the lower tail of the forecast distribution. This indicates that the forecast distribution may be overestimating the actual outcome. The closer the PIT value is to 0, the more significant the overestimation.
    \item \textbf{PIT $>$ 0.5:} Conversely, if the PIT value is greater than 0.5, the observed value lies in the upper tail of the forecast distribution, suggesting that the forecast might be underestimating the actual outcome. A PIT value closer to 1 indicates a more severe underestimation.
\end{itemize}
Comparing the PIT statistics for 2015 and 2020 for QR+PP signals for 2015, the right quantile had a PIT value of 0.43, while for 2020, it returned a PIT value of 0.86, signalling a clear underestimation. Comparing, instead, PIT values in 2020 between the QR and QR+PP, we obtained a reduction from  0.95 to 0.86, signalling an overall better model calibration by the inclusion of the PP predictor. The predicted quantiles distribution such as the one depicted in Figure \ref{fig:quantiles_dis}) could be used for placing more informed bets on a market, such as the Agora (see Section \ref{sec1:introduction}). Yet imperfect, this can represent an early integration of a risk assessment evaluation inside a specific environmental forecast; it clearly offers a more informed decision-making approach.

\section{Conclusions}
\label{sec:final}

In this paper, we address the critical challenge of forecasting the yearly number of hurricanes in the Atlantic Ocean. Understanding the factors driving this increase is essential for effective risk management, infrastructure assessment, and human vulnerability mitigation. Prior studies have delved into various aspects of hurricane forecasting, focusing predominantly on predicting hurricane paths rather than yearly counts. The literature highlights the significance of accurate hurricane count forecasts for risk assessment, emergency response coordination, and economic sectors such as manufacturing and finance. Traditional statistical methods like ARIMA models forecast combination techniques and the use of Poisson distribution have been extensively explored alongside recent advancements in deep learning techniques, particularly Recurrent Neural Networks with a Long Short-Term Memory architecture. 

Our paper contributes three key innovations. First, we employ a probabilistic forecasting approach, enhancing the traditional deterministic forecasts. Second, we introduce a novel index construction tailored specifically for assessing Atlantic hurricane activity, enhancing forecast accuracy. Third, we systematically compared different statistical models, evaluating their efficacy across various performance metrics, including absolute accuracy, quantile accuracy, and directional accuracy. Our focus on directional accuracy is particularly novel in the environmental sciences, aiming to improve the predictive ability of models by considering the direction of change in hurricane counts. Our analysis underscores the importance of simpler modelling approaches for enhancing interpretability and sensitivity, which are crucial for effective decision-making in climate-related risk scenarios. By systematically gathering data and selecting relevant predictors, we demonstrate the potential for improved forecast accuracy while ensuring reproducibility and transparency. We advocate for the adoption of simple quantile forecasts to account for potential underestimation or overestimation in decision-making processes. Yet other models like the ARIMA and INGARCH could still be considered for future development since their MAE performances resulted in being equivalent to those of QR.

Future research directions could include further exploration of probabilistic forecasting approaches, refinement of index constructions for hurricane activity assessment, and continued evaluation of different statistical models. For instance, predictors could be included with standalone indicator variables or through interaction terms. If there is a suspicion that the effects of existing predictors vary under specific conditions, it may be beneficial to create interaction terms between these predictors and a newly introduced indicator variable. Additionally, there is a need to enhance the incorporation of uncertainty evaluations in forecasting methodologies and to explore the integration of past experiences and observational evidence for improved forecast accuracy.



\section*{Acknowledgments}
For the purpose of open access, the author(s) has applied a Creative Commons Attribution (CC BY) licence to any Author Accepted Manuscript version arising from this submission.

\section*{Competing interests}
No competing interest is declared.


\bibliographystyle{plainnat}  
\bibliography{bibliography}

\begin{thebibliography}{50}
\providecommand{\natexlab}[1]{#1}
\providecommand{\url}[1]{\texttt{#1}}
\expandafter\ifx\csname urlstyle\endcsname\relax
  \providecommand{\doi}[1]{doi: #1}\else
  \providecommand{\doi}{doi: \begingroup \urlstyle{rm}\Url}\fi

\bibitem[psl()]{psl_enso_dashboard}
{NOAA} {ENSO} dashboard.
\newblock \url{https://psl.noaa.gov/enso/dashboard.html}.
\newblock Accessed: April 19, 2024.

\bibitem[Bell and Chelliah(2006)]{bell2006leading}
Gerald~D Bell and Muthuvel Chelliah.
\newblock Leading tropical modes associated with interannual and multidecadal fluctuations in north atlantic hurricane activity.
\newblock \emph{Journal of Climate}, 19\penalty0 (4):\penalty0 590--612, 2006.

\bibitem[Blaskowitz and Herwartz(2011)]{blaskowitz2011economic}
Oliver Blaskowitz and Helmut Herwartz.
\newblock On economic evaluation of directional forecasts.
\newblock \emph{International journal of forecasting}, 27\penalty0 (4):\penalty0 1058--1065, 2011.

\bibitem[Bose et~al.(2022)Bose, Pintar, and Simiu]{bose2022real}
Rikhi Bose, Adam Pintar, and Emil Simiu.
\newblock A real time prediction methodology for hurricane evolution using {LSTM} recurrent neural networks.
\newblock \emph{Neural Computing and Applications}, 34\penalty0 (20):\penalty0 17491--17505, 2022.

\bibitem[Browell and Gilbert(2020)]{browell2020probcast}
Jethro Browell and Ciaran Gilbert.
\newblock Probcast: Open-source production, evaluation and visualisation of probabilistic forecasts.
\newblock In \emph{2020 International Conference on Probabilistic Methods Applied to Power Systems (PMAPS)}, pages 1--6. IEEE, 2020.

\bibitem[Burn and Palmer(2015)]{burn2015atlantic}
Michael~J Burn and Suzanne~E Palmer.
\newblock Atlantic hurricane activity during the last millennium.
\newblock \emph{Scientific reports}, 5\penalty0 (1):\penalty0 12838, 2015.

\bibitem[Cade and Noon(2003)]{cade2003gentle}
Brian~S Cade and Barry~R Noon.
\newblock A gentle introduction to quantile regression for ecologists.
\newblock \emph{Frontiers in Ecology and the Environment}, 1\penalty0 (8):\penalty0 412--420, 2003.

\bibitem[Chang et~al.(2018)Chang, Yang, and Yu]{chang2018hurricane}
Chia-Chien Chang, Jen-Wei Yang, and Min-Teh Yu.
\newblock Hurricane risk management with climate and co2 indices.
\newblock \emph{Journal of Risk and Insurance}, 85\penalty0 (3):\penalty0 695--720, 2018.

\bibitem[Chen et~al.(2020)Chen, Zhang, and Wang]{chen2020machine}
Rui Chen, Weimin Zhang, and Xiang Wang.
\newblock Machine learning in tropical cyclone forecast modeling: A review.
\newblock \emph{Atmosphere}, 11\penalty0 (7):\penalty0 676, 2020.

\bibitem[Chylek and Lesins(2008)]{chylek2008multidecadal}
Petr Chylek and Glen Lesins.
\newblock Multidecadal variability of atlantic hurricane activity: 1851--2007.
\newblock \emph{Journal of Geophysical Research: Atmospheres}, 113\penalty0 (D22), 2008.

\bibitem[Cui and Wu(2016)]{cui2016conditional}
Yunwei Cui and Rongning Wu.
\newblock On conditional maximum likelihood estimation for ingarch (p, q) models.
\newblock \emph{Statistics \& Probability Letters}, 118:\penalty0 1--7, 2016.

\bibitem[Curtis and Adler(2003)]{curtis2003evolution}
Scott Curtis and Robert~F Adler.
\newblock Evolution of el ni{\~n}o-precipitation relationships from satellites and gauges.
\newblock \emph{Journal of Geophysical Research: Atmospheres}, 108\penalty0 (D4), 2003.

\bibitem[Daneshvaran and Haji(2012)]{daneshvaran2012atlantic}
Siamak Daneshvaran and Maryam Haji.
\newblock Atlantic hurricane forecast: a statistical analysis.
\newblock \emph{The Journal of Risk Finance}, 14\penalty0 (1):\penalty0 4--19, 2012.

\bibitem[Elsner(2006)]{elsner2006evidence}
James~B Elsner.
\newblock Evidence in support of the climate change--atlantic hurricane hypothesis.
\newblock \emph{Geophysical Research Letters}, 33\penalty0 (16), 2006.

\bibitem[Elsner et~al.(2001)Elsner, Bossak, and Niu]{elsner2001secular}
James~B Elsner, Brian~H Bossak, and Xu-Feng Niu.
\newblock Secular changes to the enso-us hurricane relationship.
\newblock \emph{Geophysical Research Letters}, 28\penalty0 (21):\penalty0 4123--4126, 2001.

\bibitem[Elsner et~al.(2008)Elsner, Jagger, Dickinson, and Rowe]{elsner2008improving}
James~B Elsner, Thomas~H Jagger, Michael Dickinson, and Dail Rowe.
\newblock Improving multiseason forecasts of north atlantic hurricane activity.
\newblock \emph{Journal of Climate}, 21\penalty0 (6):\penalty0 1209--1219, 2008.

\bibitem[Elsner et~al.(1998)Elsner, Niu, and Tsonis]{elsner1998multi}
JB~Elsner, X~Niu, and AA~Tsonis.
\newblock Multi-year prediction model of north atlantic hurricane activity.
\newblock \emph{Meteorology and Atmospheric Physics}, 68:\penalty0 43--51, 1998.

\bibitem[Goldenberg et~al.(2001)Goldenberg, Landsea, Mestas-Nu{\~n}ez, and Gray]{goldenberg2001recent}
Stanley~B Goldenberg, Christopher~W Landsea, Alberto~M Mestas-Nu{\~n}ez, and William~M Gray.
\newblock The recent increase in atlantic hurricane activity: Causes and implications.
\newblock \emph{Science}, 293\penalty0 (5529):\penalty0 474--479, 2001.

\bibitem[Granger and Pesaran(2000)]{granger2000economic}
Clive~WJ Granger and M~Hashem Pesaran.
\newblock Economic and statistical measures of forecast accuracy.
\newblock \emph{Journal of Forecasting}, 19\penalty0 (7):\penalty0 537--560, 2000.

\bibitem[Hyndman and Khandakar(2008)]{hyndman2008automatic}
Rob~J Hyndman and Yeasmin Khandakar.
\newblock Automatic time series forecasting: the forecast package for r.
\newblock \emph{Journal of statistical software}, 27:\penalty0 1--22, 2008.

\bibitem[Iman et~al.(2006)Iman, Johnson, and Watson~Jr]{iman2006statistical}
Ronald~L Iman, Mark~E Johnson, and Charles~C Watson~Jr.
\newblock Statistical aspects of forecasting and planning for hurricanes.
\newblock \emph{The American Statistician}, 60\penalty0 (2):\penalty0 105--121, 2006.

\bibitem[Jagger and Elsner(2010)]{jagger2010consensus}
Thomas~H Jagger and James~B Elsner.
\newblock A consensus model for seasonal hurricane prediction.
\newblock \emph{Journal of Climate}, 23\penalty0 (22):\penalty0 6090--6099, 2010.

\bibitem[Keenan(2019)]{keenan2019climate}
Jesse~M Keenan.
\newblock A climate intelligence arms race in financial markets.
\newblock \emph{Science}, 365\penalty0 (6459):\penalty0 1240--1243, 2019.

\bibitem[Klotzbach et~al.(2017)Klotzbach, Saunders, Bell, and Blake]{klotzbach2017north}
Philip~J Klotzbach, Mark~A Saunders, Gerald~D Bell, and Eric~S Blake.
\newblock North atlantic seasonal hurricane prediction: underlying science and an evaluation of statistical models.
\newblock \emph{Climate extremes: Patterns and mechanisms}, pages 315--328, 2017.

\bibitem[Knaff et~al.(2007)Knaff, Sampson, DeMaria, Marchok, Gross, and McAdie]{knaff2007statistical}
John~A Knaff, Charles~R Sampson, Mark DeMaria, Timothy~P Marchok, James~M Gross, and Colin~J McAdie.
\newblock Statistical tropical cyclone wind radii prediction using climatology and persistence.
\newblock \emph{Weather and Forecasting}, 22\penalty0 (4):\penalty0 781--791, 2007.

\bibitem[Koenker(2017)]{koenker2017quantile}
Roger Koenker.
\newblock Quantile regression: 40 years on.
\newblock \emph{Annual review of economics}, 9:\penalty0 155--176, 2017.

\bibitem[Kourentzes et~al.(2014)Kourentzes, Barrow, and Crone]{kourentzes2014neural}
Nikolaos Kourentzes, Devon~K Barrow, and Sven~F Crone.
\newblock Neural network ensemble operators for time series forecasting.
\newblock \emph{Expert Systems with Applications}, 41\penalty0 (9):\penalty0 4235--4244, 2014.

\bibitem[Krishnamurti et~al.(2016)Krishnamurti, Kumar, Simon, Bhardwaj, Ghosh, and Ross]{krishnamurti2016review}
TN~Krishnamurti, V~Kumar, A~Simon, A~Bhardwaj, T~Ghosh, and R~Ross.
\newblock A review of multimodel superensemble forecasting for weather, seasonal climate, and hurricanes.
\newblock \emph{Reviews of Geophysics}, 54\penalty0 (2):\penalty0 336--377, 2016.

\bibitem[Landsea and Gray(1992)]{landsea1992strong}
Christopher~W Landsea and William~M Gray.
\newblock The strong association between western sahelian monsoon rainfall and intense atlantic hurricanes.
\newblock \emph{Journal of Climate}, 5\penalty0 (5):\penalty0 435--453, 1992.

\bibitem[Letson et~al.(2007)Letson, Sutter, and Lazo]{letson2007economic}
David Letson, Daniel~S Sutter, and Jeffrey~K Lazo.
\newblock Economic value of hurricane forecasts: An overview and research needs.
\newblock \emph{Natural Hazards Review}, 8\penalty0 (3):\penalty0 78--86, 2007.

\bibitem[Livsey et~al.(2018)Livsey, Lund, Kechagias, and Pipiras]{livsey2018multivariate}
James Livsey, Robert Lund, Stefanos Kechagias, and Vladas Pipiras.
\newblock Multivariate integer-valued time series with flexible autocovariances and their application to major hurricane counts.
\newblock \emph{The Annals of Applied Statistics}, 12\penalty0 (1):\penalty0 408--431, 2018.

\bibitem[Lyubchich and Gel(2017)]{lyubchich2017can}
Vyacheslav Lyubchich and YR~Gel.
\newblock Can we weather proof our insurance?
\newblock \emph{Environmetrics}, 28\penalty0 (2):\penalty0 e2433, 2017.

\bibitem[Mann and Emanuel(2006)]{mann2006atlantic}
Michael~E Mann and Kerry~A Emanuel.
\newblock Atlantic hurricane trends linked to climate change.
\newblock \emph{Eos, Transactions American Geophysical Union}, 87\penalty0 (24):\penalty0 233--241, 2006.

\bibitem[Meyer et~al.(2014)Meyer, Horowitz, Wilks, and Horowitz]{meyer2014novel}
Robert~J Meyer, Michael Horowitz, Daniel~S Wilks, and Kenneth~A Horowitz.
\newblock A novel financial market for mitigating hurricane risk. part ii: Empirical validation.
\newblock \emph{Weather, climate, and society}, 6\penalty0 (3):\penalty0 318--330, 2014.

\bibitem[Moharana and Swain(2023)]{moharana2023recent}
Sidha~Sankalpa Moharana and Debadatta Swain.
\newblock On the recent increase in atlantic ocean hurricane activity and influencing factors.
\newblock \emph{Natural Hazards}, 118\penalty0 (2):\penalty0 1387--1399, 2023.

\bibitem[Naftaly et~al.(1997)Naftaly, Intrator, and Horn]{naftaly1997optimal}
Ury Naftaly, Nathan Intrator, and David Horn.
\newblock Optimal ensemble averaging of neural networks.
\newblock \emph{Network: Computation in Neural Systems}, 8\penalty0 (3):\penalty0 283, 1997.

\bibitem[{National Hurricane Center}(2024)]{nhc2024}
{National Hurricane Center}.
\newblock Tropical cyclone reports index, 2024.
\newblock URL \url{https://www.nhc.noaa.gov/TCR_StormReportsIndex.xml}.
\newblock Accessed: May 10, 2024.

\bibitem[{National Oceanic and Atmospheric Administration}()]{nhc_website}
{National Oceanic and Atmospheric Administration}.
\newblock {NOAA} national hurricane center.
\newblock Online.
\newblock URL \url{https://www.nhc.noaa.gov/}.

\bibitem[Parisi and Lund(2000)]{parisi2000seasonality}
Francis Parisi and Robert Lund.
\newblock Seasonality and return periods of landfalling atlantic basin hurricanes.
\newblock \emph{Australian \& New Zealand Journal of Statistics}, 42\penalty0 (3):\penalty0 271--282, 2000.

\bibitem[Philp et~al.(2019)Philp, Sabbatelli, Robertson, and Wilson]{philp2019issues}
Tom Philp, Tom Sabbatelli, Christina Robertson, and Paul Wilson.
\newblock Issues of importance to the (re) insurance industry: A timescale perspective.
\newblock \emph{Hurricane risk}, pages 1--22, 2019.

\bibitem[Qin et~al.(2021)Qin, Tang, Lu, and Lao]{qin2021trajectory}
Wanting Qin, Jun Tang, Cong Lu, and Songyang Lao.
\newblock Trajectory prediction based on long short-term memory network and {Kalman} filter using hurricanes as an example.
\newblock \emph{Computational Geosciences}, 25:\penalty0 1005--1023, 2021.

\bibitem[Russell et~al.(2020)Russell, Risser, Smith, and Kunkel]{russell2020investigating}
Brook~T Russell, Mark~D Risser, Richard~L Smith, and Kenneth~E Kunkel.
\newblock Investigating the association between late spring gulf of mexico sea surface temperatures and us gulf coast precipitation extremes with focus on hurricane harvey.
\newblock \emph{Environmetrics}, 31\penalty0 (2):\penalty0 e2595, 2020.

\bibitem[Sazcha and Christopher(2019)]{sazcha2019increases}
O~Sazcha and H~Christopher.
\newblock Increases in the extreme rainfall events: using the weibull distribution [j].
\newblock \emph{Environmetrics}, 30\penalty0 (4), 2019.

\bibitem[Shaziayani et~al.(2021)Shaziayani, Ul-Saufie, Ahmat, and Al-Jumeily]{shaziayani2021coupling}
Wan~Nur Shaziayani, Ahmad~Zia Ul-Saufie, Hasfazilah Ahmat, and Dhiya Al-Jumeily.
\newblock Coupling of quantile regression into boosted regression trees {(BRT)} technique in forecasting emission model of {PM10} concentration.
\newblock \emph{Air Quality, Atmosphere \& Health}, 14\penalty0 (10):\penalty0 1647--1663, 2021.

\bibitem[Taieb et~al.(2012)Taieb, Bontempi, Atiya, and Sorjamaa]{taieb2012review}
Souhaib~Ben Taieb, Gianluca Bontempi, Amir~F Atiya, and Antti Sorjamaa.
\newblock A review and comparison of strategies for multi-step ahead time series forecasting based on the nn5 forecasting competition.
\newblock \emph{Expert systems with applications}, 39\penalty0 (8):\penalty0 7067--7083, 2012.

\bibitem[Vasseur and Aznarte(2021)]{vasseur2021comparing}
Sebastien~P{\'e}rez Vasseur and Jos{\'e}~L Aznarte.
\newblock Comparing quantile regression methods for probabilistic forecasting of {NO2} pollution levels.
\newblock \emph{Scientific Reports}, 11\penalty0 (1):\penalty0 11592, 2021.

\bibitem[Villarini et~al.(2010)Villarini, Vecchi, and Smith]{villarini2010modeling}
Gabriele Villarini, Gabriel~A Vecchi, and James~A Smith.
\newblock Modeling the dependence of tropical storm counts in the north atlantic basin on climate indices.
\newblock \emph{Monthly Weather Review}, 138\penalty0 (7):\penalty0 2681--2705, 2010.

\bibitem[Wei{\ss}(2018)]{weiss2018introduction}
Christian~H Wei{\ss}.
\newblock \emph{An introduction to discrete-valued time series}.
\newblock John Wiley \& Sons, 2018.

\bibitem[WMO(1993)]{world1993global}
WMO.
\newblock \emph{Global guide to tropical cyclone forecasting}.
\newblock Secretariat of the World Meteorological Organization, 1993.

\bibitem[Xiao et~al.(2015)Xiao, Kottas, and Sans{\'o}]{xiao2015modeling}
Sai Xiao, Athanasios Kottas, and Bruno Sans{\'o}.
\newblock Modeling for seasonal marked point processes: An analysis of evolving hurricane occurrences.
\newblock \emph{The Annals of Applied Statistics}, pages 353--382, 2015.

\end{thebibliography}

\newpage

\appendix

\end{document}